# On the Hardness of Welfare Maximization in Combinatorial Auctions with Submodular Valuations

Shahar Dobzinski      Jan Vondrák

August 8, 2018

**Abstract**

We present a new type of monotone submodular functions: *multi-peak submodular functions*. Roughly speaking, given a family of sets $\mathcal{F}$, we construct a monotone submodular function $f$ with a high value $f(S)$ for every set $S \in \mathcal{F}$ (a "peak"), and a low value on every set that does not intersect significantly any set in $\mathcal{F}$.

We use this construction to show that a better than $(1 - \frac{1}{2e})$-approximation ($\simeq 0.816$) for welfare maximization in combinatorial auctions with submodular valuations is (1) impossible in the communication model, (2) NP-hard in the computational model where valuations are given explicitly. Establishing a constant approximation hardness for this problem in the communication model was a long-standing open question. The valuations we construct for the hardness result in the computational model depend only on a constant number of items, and hence the result holds even if the players can answer arbitrary queries about their valuation, including demand queries.

We also study two other related problems that received some attention recently: max-min allocation (for which we also get hardness of $(1 - \frac{1}{2e} + \epsilon)$-approximation, in both models), and combinatorial public projects (for which we prove hardness of $(\frac{3}{4} + \epsilon)$-approximation in the communication model, and hardness of $(1 - \frac{1}{e} + \epsilon)$-approximation in the computational model, using constant size valuations).

## 1 Introduction

In a combinatorial auction there are $k$ players and a set of $M$ items ($|M| = m$). Each players $i$ has a valuation function $v_i : 2^M \to R_+$. In this paper we use the standard assumption that the valuation functions are non-decreasing and that $v_i(\emptyset) = 0$. The goal is to find an allocation of the items $(S_1, \ldots, S_k)$ to maximize the social welfare $\Sigma_i v_i(S_i)$. There is extensive literature on combinatorial auctions, and the two most popular lines of research in this area are the design of computationally efficient truthful mechanisms for combinatorial auctions and the development of (non-truthful) approximation algorithms (for various restricted classes of valuations, such as subadditive [13, 10] or submodular [21, 16]). This paper belongs to the latter type.

Specifically, we consider the case where the valuation functions are submodular, and ask what is the best approximation of the optimum social welfare that is achievable in polynomial time. The case of players with submodular valuations unique because of two reasons. First, submodular valuations have a natural economic interpretation of decreasing marginal utilities. Second, the rich literature on submodular optimization makes the study of combinatorial auctions with submodular valuations an ideal midway point for the exchange of techniques and ideas between algorithmic game theory and combinatorial optimization.

We would like our algorithms to run in time that is polynomial in $k$ and $m$. However, in a naive representation of the valuation function we may have to specify one number for each possible bundle, so the total size is exponential in $m$. Algorithms for combinatorial auctions usually assume that the valuations are represented by oracles. The algorithm is assumed to make only a polynomial number of queries to the oracles. Popular queries include the value query (given a set $S$, what is $v(S)$?) and the stronger demand oracle (given prices $p_1, \ldots, p_m$ return a bundle $S \in \arg\max v(S) - \Sigma_{i \in S} p_i$). The state of the art is an $(1 - 1/e)$-approximation that uses value queries only [26] (which is optimal in the value oracle model [19, 22]), and a $(1-1/e+\delta)$-approximation that uses demand queries only [17] (where $\delta > 0$ is some small constant). As one can observe from these results, different oracles may potentially enable us to obtain different approximation ratios. Our goal in this paper is to prove hardness results that will hold for *every* oracle.

The literature contains two approaches to deal with this issue:

- **Communication complexity:** Nisan and Segal pioneered this approach [24, 25]. Players can transmit any information and thus answer any possible oracle query, but the number of bits transmitted is polynomially bounded. The algorithm and the players are conservatively assumed to be computationally unbounded.

- **Constant-size valuations:** This approach was introduced in [17]. Each player has a valuation function that depends non-trivially only on a constant number of items, and its full description is available to the algorithm (so any query whose answer depends only on the valuation can be answered in constant time – see a discussion in [16]). The algorithm is required to run in polynomial time.

The two approaches are conceptually at two opposite ends of the spectrum: either arbitrarily complex valuations and infinite computational power, limited only by the communication bottleneck, or extremely succinct valuations and a polynomial-time algorithm.

## 1.1 Our Main Result: The Hardness of Welfare Maximization

We prove the following hardness result for welfare maximization, with the same factor in both models.

**Theorem 1.1.** *For welfare maximization with submodular valuations, a $(1 - \frac{1}{2e} + \epsilon)$-approximation for any constant $\epsilon > 0$ requires exponential communication. For constant-size submodular valuations, a $(1 - \frac{1}{2e} + \epsilon)$-approximation would imply $P = NP$.*

We note that previously, only $(1 - o(1))$-hardness was known in the communication complexity model, and we show the first constant-factor hardness in this model (this question was open since the seminal paper of Nisan and Segal [25]). In particular, this settles affirmatively a conjecture of [10]. NP-hardness of $(1-\rho)$-approximation for some very small $\rho > 0$ was known for constant-size submodular valuations [16].

Each of our results implies hardness of $(1-\frac{1}{2e}+\epsilon)$-approximation in the demand oracle model, because demand queries can be simulated in polynomial communication, and because demand queries can be efficiently implemented for constant-size valuations. This improves the known NP-hardness of $(\frac{15}{16}+\epsilon)$-approximation in the demand oracle model by Chakrabarty and Goel [8]. We stress that our results apply to any "reasonable" oracle model and not just to the demand oracle model; see also the discussion in [16].

We remark that the communication complexity result holds even for a constant number of players, in the sense that for every $\epsilon > 0$ there is some fixed $k \geq 2$ such that the hardness result holds for $k$ players. For the special case of 2 players, we get the following.

**Theorem 1.2.** *For welfare maximization with 2 players with submodular valuations, a $(\frac{17}{18}+\epsilon)$-approximation for any $\epsilon > 0$ requires exponential communication.*

For constant-size valuations, we clearly need a superconstant number of players, otherwise the problem can be solved in constant time.

## 1.2 Our Techniques

At the heart of our hardness proofs is a new construction of submodular functions, which has been the stumbling block of previous attempts to prove similar results. We call these *multiple-peak submodular functions*: given a family of (possibly overlapping, exponentially many) sets $\mathcal{F}$, we want a monotone submodular function with high value $f(S)$ for every set $S \in \mathcal{F}$ (a "peak"), and a low value on every set that does not intersect significantly any set in $\mathcal{F}$. The precise parameters depend on the family $\mathcal{F}$, in particular how much the sets in $\mathcal{F}$ are allowed to overlap. As an aside, this construction goes in the opposite direction to the (very different) construction of Balcan and Harvey [4], where a submodular function is constructed that has very *low value* on a given family of sets.

| Problem | Approximation | CC hardness (2 players) | CC hardness | NP-hardness |
|---|---|---|---|---|
| submodular welfare | $1 - 1/e + \delta$ | $17/18$ | $1 - 1/(2e)$ | $1 - 1/(2e)$ |
| submodular max-min* | $1 - 1/e$ | $17/18$ | $1 - 1/(2e)$ | $1 - 1/(2e)$ |
| submodular CPP | $1 - 1/e$ | $7/8$ | $3/4$ | $1 - 1/e$ |

Figure 1: Summary of our hardness results: "CC hardness" refers to communication complexity, or more precisely inapproximability under polynomial communication, and "NP-hardness" refers to inapproximability for constant-size valuation functions. The approximation results shown here are in the demand oracle model for submodular welfare [17], in the value oracle model for submodular max-min allocation (*for a constant number of players) [9], and in the value oracle model for submodular CPP [23]. In the value oracle model, $1 - 1/e$ is optimal for all 3 problems.

This construction is then leveraged in two possible ways: for communication complexity, we use the general framework introduced by Nisan [24]. We use the probabilistic method to prove the existence of an exponentially large family $\mathcal{F}$ and show a reduction from the Set Disjointness problem. Each player has a multiple-peak submodular valuation corresponding to a family $\mathcal{F}_i$, where either there is a choice of sets $S_i \in \mathcal{F}_i$ that are "consistent" (disjoint/equal, depending on the problem), or not - distinguishing these two cases requires exponential communication.

For NP-hardness results, we start from Feige's proof of inapproximability of Max $k$-cover [14]. Then we use the coverage system to construct an instance with multi-peak submodular functions. In the YES case, the instance contains matching sets, providing high value for each player, while in the NO case, some players are forced to receive low value. In addition, Feige's construction guarantees that the valuation functions depend only on a constant (albeit large) number of items. Hence the computational hardness result applies even if the bidders have constant-size valuations.

## 1.3 Other Results

We use the new multiple-peak submodular functions to prove hardness results for every oracle in two other related settings. One is Combinatorial Public Projects (CPP), where we seek a single set $S$ of at most $k$ items, maximizing $\sum_{i=1}^{n} v_i(S)$. Another one is the Max-min Allocation Problem, where we want to allocated disjoint sets of items $S_i$ in order to maximize $\min_{1 \leq i \leq k} v_i(S_i)$.

**Combinatorial Public Projects.** For CPP there is an $(1 - 1/e)$-approximation that can be achieved using value queries only [23]. We show that this is optimal in any arbitrarily powerful oracle model even for constant size valuation, hence strengthening previous result that showed that this result cannot be improved using demand queries [6]. We note that the situation is different for the welfare maximization, where $1 - 1/e$ is optimal for value queries but not optimal if demand queries are allowed [17]. No constant-factor communication complexity hardness was previously known for the CPP problem.

**Theorem 1.3.** *For the CPP problem with submodular valuations, a $(\frac{3}{4} + \epsilon)$-approximation for any $\epsilon > 0$ requires exponential communication. For 2 players with submodular valuations, a $(\frac{7}{8} + \epsilon)$-approximation for any $\epsilon > 0$ requires exponential communication. For constant-size submodular valuations, a $(1 - \frac{1}{e} + \epsilon)$-approximation would imply $P = NP$.*

**Max-min Allocation.** Finally, our results for Welfare Maximization also apply to the Submodular Max-min Allocation problem. The problem of obtaining a max-min allocation was quite heavily studied recently [5, 3, 12, 2, 7], and most of the work concentrated on cases where the valuations are special cases of submodular valuations. It is known that the Submodular Max-min Allocation problem admits a $(1 - 1/e - \epsilon)$-approximation for any fixed $\epsilon > 0$ and any fixed number of players $k$ [9], and this is optimal in the value oracle model, if we want approximation independent of $k$ [22]. Our hardness results also hold for a fixed number of players $k$ (approaching $1 - \frac{1}{2e}$ for $k$ fixed but large), and they are independent of a particular oracle model. When the number of players is not fixed, the problem is less understood, and the best known approximation is $O(m^{1/2+\epsilon})$ [18, 7].

**Theorem 1.4.** *For Max-min Allocation with submodular valuations, a $(1 - \frac{1}{2e} + \epsilon)$-approximation for any $\epsilon > 0$ requires exponential communication. For constant-size submodular valuations, a $(1 - \frac{1}{2e} + \epsilon)$-approximation would imply $P = NP$. For Max-min Allocation with 2 players with submodular valuations, a $(\frac{17}{18} + \epsilon)$-approximation for any $\epsilon > 0$ requires exponential communication.*

The proof is obtained by a straightforward adaption of the proof for Welfare Maximization, as we explain in Section 4.3.

**Organization.** The rest of the paper is organized as follows. In Section 2, we give the necessary preliminaries. In Section 3, we present the construction of multiple-peak submodular functions, the main technical ingredient of our proofs. In Section 4, we present the proofs for Welfare Maximization, and we also note why they apply to Max-min Allocation. In Section B, we present the proofs for Combinatorial Public Projects.

## 2 Preliminaries

A function $f : \{0,1\}^n \to \mathbb{R}$ is submodular if $f(A) + f(B) \geq f(A \cup B) + f(A \cap B)$ for all $A, B$ (and sets are naturally identified with $\{0,1\}$ vectors). By calling a function monotone, we mean non-decreasing, i.e. $f(A) \leq f(B)$ for all $A \subseteq B$. We use the following connection between continuous and discrete submodular functions (slightly modifying the conditions used in [20] and [27], where "smooth submodular functions" were used).

**Definition 2.1.** *We call a function $F : [0,1]^n \to \mathbb{R}$ continuous submodular, if all of the following conditions hold:*

- *$F$ is absolutely continuous[1] on every line segment of direction $\mathbf{e}_i$, for every $i$.*
- *The partial derivative $\frac{\partial F}{\partial x_i}$ is defined almost everywhere on every line segment of direction $\mathbf{e}_i$.*
- *$\frac{\partial F}{\partial x_i}$ is non-increasing with respect to every coordinate.*

**Lemma 2.2.** *If $F : [0,1]^n \to \mathbb{R}$ is continuous submodular, then its restriction to $\{0,1\}^n$ is submodular.*

*Proof.* Since $F$ is absolutely continuous, we can express differences between point values by integrating the partial derivatives: assuming that $\mathbf{x} \in \{0,1\}^n$ and $x_i = x_j = 0$,

$$F(\mathbf{x} + \mathbf{e}_i) - F(\mathbf{x}) = \int_0^1 \frac{\partial F}{\partial x_i}\bigg|_{\mathbf{x}+t\mathbf{e}_i} dt.$$

Since $\frac{\partial F}{\partial x_i}$ is non-increasing with respect to every coordinate, in particular $x_j$, we get

$$F(\mathbf{x} + \mathbf{e}_i + \mathbf{e}_j) - F(\mathbf{x} + \mathbf{e}_j) \leq F(\mathbf{x} + \mathbf{e}_i) - F(\mathbf{x}).$$

This means that $F$ restricted to $\{0,1\}^n$ is submodular. □

In this paper, $(x)_+ = \max\{x, 0\}$ denotes the positive part of a number. We remark that when we write $(x)_+^2$, we mean $(\max\{x, 0\})^2$, i.e. for $x < 0$ this quantity is 0.

---
[1] A function $F : \mathbb{R} \to \mathbb{R}$ is absolutely continuous if for every $\epsilon > 0$ there is $\delta > 0$ such that whenever $\sum_i |x_i - y_i| < \delta$, $\sum_i |F(x_i) - F(y_i)| < \epsilon$.

# 3 Construction of Multiple-Peak Submodular Valuations

In this section, we construct a new class of submodular functions. These functions are the main technical elements of our hardness results. Given a family of sets $\mathcal{F}$, we would like to define a function that has a high value on each set in $\mathcal{F}$, and low value on every set that does not overlap any set in $\mathcal{F}$ very much. We call these functions multiple-peak submodular valuations.

The construction is inspired by the "continuous point of view", which has been used in previous work on submodular optimization [22, 26, 27], and the connection between continuous and discrete submodular functions given by Lemma 2.2. Suppose we start from a continuous submodular function $F : [0,1]^M \to \mathbb{R}_+$ that depends only on the sum of coordinates $\sum x_i$, in particular $F(\mathbf{x}) = 1 - (1 - a\sum x_i)_+^2$. In addition, we have a collection $\mathcal{F}$ of vertices of the hypercube that should become the "peaks" of the function, i.e. their value should be increased compared to $F(\mathbf{x})$. We define a region of the hypercube $B_A \subset [0,1]^M$ around each vertex $A \in \mathcal{F}$, such that these regions are disjoint. Then if we modify the function $F$ so that it is continuous submodular on each region separately, and first partial derivatives are continuous on the boundary of each region, then the resulting function is still continuous submodular. Crucially, we can perform this operation on each region $B_A$ independently, since there is no interaction between different (disjoint) regions, due to the local nature of continuous submodularity - we only have to ensure that first partial derivatives are continuous on the boundary of each region.

In particular, consider a region defined by $B_A = \{\mathbf{x} \in [0,1]^M : \sum_{i \in A} x_i - \sum_{i \notin A} x_i > b\}$. We can define $F_A(\mathbf{x}) = 1 - (1 - a(2\sum_{i \in A} x_i - b))_+(1 - a(2\sum_{i \notin A} x_i + b))_+$. This function is continuous submodular on $B_A$. Moreover, it agrees with $F(\mathbf{x})$ on the boundary of $B_A$ and also its first partial derivatives are the same. Therefore, we can glue the two functions seamlessly at the boundary of $B_A$ and the resulting function is still continuous submodular.

A discrete formulation of this idea follows.

**Definition 3.1.** *A set $S \subseteq M$ is said to be $b$-close to a set $A$, if $|S \cap A| - |S \setminus A| > b$. More generally, a point $\mathbf{x} \in [0,1]^M$ is $b$-close to $A$, if $\sum_{i \in A} x_i - \sum_{i \notin A} x_i > b$.*

Note that $S$ being $b$-close to $A$ is the same as the characteristic vector $\mathbf{1}_S$ being $b$-close to $A$.

**Proposition 3.2.** *Let $A, A' \subseteq M$ be two sets such that $|A \cap A'| \leq b$. Then every point $\mathbf{x} \in [0,1]^M$ is $b$-close to at most one of the two sets $A, A'$. In particular, every set $S \subseteq M$ is $b$-close to at most one of $A, A'$.*

*Proof.* Suppose towards contradiction that $\mathbf{x}$ is $b$-close to both $A$ and $A'$. That is, $\sum_{i \in A} x_i - \sum_{i \notin A} x_i > b$ and $\sum_{i \in A'} x_i - \sum_{i \notin A'} x_i > b$. Adding up these two inequalities, we get

$$2b < (\sum_{i \in A} x_i - \sum_{i \notin A} x_i) + (\sum_{i \in A'} x_i - \sum_{i \notin A'} x_i)$$

$$= (\sum_{i \in A} x_i - \sum_{i \notin A} x_i + \sum_{i \in A} x_i - \sum_{i \in A} x_i) + (\sum_{i \in A'} x_i - \sum_{i \notin A'} x_i + \sum_{i \in A'} x_i - \sum_{i \in A'} x_i)$$

$$= (2\sum_{i \in A} x_i - \sum_{i} x_i) + (2\sum_{i \in A} x_i - \sum_{i} x_i)$$

$$= 2\sum_{i \in A} x_i + 2\sum_{i \in A'} x_i - 2\sum_{i} x_i$$

$$= 2\sum_{i \in A \cap A'} x_i + 2\sum_{i \in A \cup A'} x_i - 2\sum_{i} x_i \leq 2|A \cap A'| \leq 2b$$

We have reached a contradiction and therefore $\mathbf{x}$ is close to at most one of the sets. $\square$

**Definition 3.3.** *A family of sets $\mathcal{F} \subset 2^M$ is called $b$-intersecting if $|A \cap A'| \leq b$ for all $A, A' \in \mathcal{F}, A \neq A'$.*

**Definition 3.4** (Multiple-Peak Submodular Valuation). *Let $\mathcal{F}$ be a $b$-intersecting family and $a > 0$ some constant. A function $f$ is called $(\mathcal{F}, a, b)$-multiple-peak function if it is defined as follows:*

1. For a set $S$ that is $b$-close to some $A \in \mathcal{F}$ (unique by the above),
$$f(S) = 1 - (1 - a(2|S \cap A| - b))_+ (1 - a(2|S \setminus A| + b))_+.$$

2. For a set $S$ that is not $b$-close to any $A \in \mathcal{F}$,
$$f(S) = 1 - (1 - a|S|)_+^2.$$

**Lemma 3.5.** *For every constant $a > 0$ and every $b$-intersecting family $\mathcal{F}$, the $(\mathcal{F}, a, b)$-multiple-peak function $f$ (as defined in Definition 3.4) is monotone and submodular.*

*Proof.* We define the following function $F : [0, 1]^M \to \mathbb{R}_+$:
$$F(\mathbf{x}) = 1 - (1 - a \sum_i x_i)_+^2.$$

One can verify that this function is continuous submodular, because it is absolutely continuous and its partial derivatives are non-increasing. In addition, for each $A \in \mathcal{F}$, we define
$$F_A(\mathbf{x}) = 1 - (1 - a(2 \sum_{i \in A} x_i - b))_+ (1 - a(2 \sum_{i \notin A} x_i + b))_+$$

Again, it is easy to see that $F_A$ is continuous submodular, by checking that its partial derivatives are non-increasing.

Define $B_A = \{\mathbf{x} \in [0, 1]^M : \sum_{i \in A} x_i - \sum_{i \notin A} x_i > b\}$, i.e. the region $b$-close to $A$. We know that these regions are disjoint for $A \in \mathcal{F}$. We define a function $\tilde{F}(\mathbf{x})$ to be equal to $F_A(\mathbf{x})$ for $\mathbf{x} \in B_A$, and equal to $F(\mathbf{x})$ when $\mathbf{x} \notin B_A$ for any $A \in \mathcal{F}$.

We need to prove that $\tilde{F}$ is a continuous submodular function. Since $F$ and $F_A$ are continuous submodular functions, we only have to verify that they agree on the boundary of $B_A$ up to first partial derivatives. Note that the boundary of $B_A$ intersects any axis-parallel line segment in at most one point. Therefore this will imply that on any line segment parallel to $\mathbf{e}_i$, $\tilde{F}$ is absolutely continuous, $\frac{\partial \tilde{F}}{\partial x_i}$ is defined almost everywhere, and it is non-increasing with respect to every coordinate. Therefore, $\tilde{F}$ is continuous submodular.

For each boundary point $\mathbf{x}$ of $B_A$, we have $2 \sum_{i \in A} x_i - b = 2 \sum_{i \notin A} x_i + b = \sum_i x_i$. Therefore, $F_A(\mathbf{x}) = F(\mathbf{x})$. In addition, if $1 - a \sum_i x_i \leq 0$, then $F(\mathbf{x}) = F_A(\mathbf{x}) = 1$ and all partial derivatives are 0. Finally, assume that $1 - a \sum_i x_i > 0$. For $j \in A$ we have
$$\frac{\partial F_A}{\partial x_j} = 2a(1 - a(2 \sum_{i \notin A} x_i + b)) = 2a(1 - a \sum_i x_i) = \frac{\partial F}{\partial x_j}.$$

Similarly, for $j \notin A$, we have
$$\frac{\partial F_A}{\partial x_j} = 2a(1 - a(2 \sum_{i \in A} x_i - b)) = 2a(1 - a \sum_i x_i) = \frac{\partial F}{\partial x_j}.$$

Hence $F$ and $F_A$ agree on the boundary of $B_A$ up to their first partial derivatives.

Therefore, $\tilde{F}$ is a continuous submodular function and also monotone (non-decreasing). By Lemma 2.2, its restriction to $\{0, 1\}^M$ is monotone submodular. This is the multi-peak submodular function of Definition 3.4. □

## 4 Welfare Maximization

In this section we prove our three results regarding welfare maximization: for every constant $\epsilon > 0$, a $(1 - \frac{1}{2e} + \epsilon)$-approximation requires exponential communication, for 2 players, a $\frac{17}{18} + \epsilon$ requires exponential communication and that for constant-size submodular valuations, a $(1 - \frac{1}{2e} + \epsilon)$-approximation would imply $P = NP$. We then note why the same proof implies hardness result for max-min allocations (Theorem 1.4). All results share the same basic ideas and use multi-peak submodular functions. The multi-peak submodular functions are defined using a special kind of a set system:

**Definition 4.1.** *A collection of sets $\mathcal{S} \subset 2^M$ that is partitioned into groups $\mathcal{S}_1, \ldots, \mathcal{S}_k$ is called* well structured *if all of the following holds:*

- *The total number of elements in the universe is $m = k \cdot s$.*
- *There is some number $s$ such that for each set $S \in \mathcal{S}$, $|S| = s$.*
- *$|\mathcal{S}_1| = \ldots = |\mathcal{S}_k|$.*
- *There exists some $b$ such that for every $i \in [k]$ and every $S, S' \in \mathcal{S}_i$, we have $|S \cap S'| \leq b$.*

Given a well structured collection, we identify the $k$ groups of sets with $k$ players. The set of items is $M$. For each player $i$, we define a $(\mathcal{S}_i, a, b)$-multi-peak submodular valuation function $v_i : 2^M \to \mathbb{R}_+$. The precise choice of $a, b$ is application specific and will be given later.

We will say that an instance is a *YES* instance if there exist $k$ disjoint sets, one from each group $\mathcal{S}_i$, whose union is $M$. An instance is a *NO* instance if for any choice of $\ell \leq k$ sets, their union covers at most a $(1 - (1 - 1/k)^\ell + \epsilon)m$ items.

The inapproximability gap that we prove is the result of the intractability of distinguishing between YES instances and NO instances. We start with calculating the welfare in YES instances. The more involved task is to upper bound the welfare in NO instances.

**Claim 4.2.** *The optimal welfare in a YES instance is at least $k(1 - (1 - a(2s - b))_+ (1 - ab)_+)$.*

*Proof.* This is a YES instance so for each bidder $i$ there is a set $S_i \in \mathcal{S}_i$ such that the $S_i$'s are disjoint and cover the universe. A feasible allocation gives $S_i$ to player $i$, and each player collects value $v_i(S_i) = 1 - (1 - a(2s - b))_+ (1 - ab)_+$. So the overall welfare of this allocation is $k(1 - (1 - a(2s - b))_+ (1 - ab)_+)$. □

Consider now a NO instance. There is no such choice of disjoint sets as in YES instances, and moreover any selection of $\ell$ sets such that $A_i \in \mathcal{S}_i$ satisfies $|\bigcup_{i=1}^\ell A_i| \leq (1 - (1 - 1/k)^\ell + \epsilon)ks$. Consider any feasible allocation of disjoint sets $(S_1, \ldots, S_k)$. For each $S_i$, let $A_i \in \mathcal{S}_i$ be the set in $\mathcal{S}_i$ closest to $S_i$, i.e. maximizing $|S_i \cap A_i| - |S_i \setminus A_i|$. Our goal now is to upper bound the maximum welfare such allocation can achieve.

For each $i$, we define variables $x_i = \frac{1}{s}|S_i \cap A_i|, y_i = \frac{1}{s}|S_i \setminus A_i|$. Without loss of generality, $x_1 \geq x_2 \geq x_3 \geq \ldots \geq x_k$. For every $\ell$, the first $\ell$ variables $x_i$ satisfy the constraint that $\sum_{i=1}^\ell x_i \leq (1 - (1 - 1/k)^\ell + \epsilon)k$ (arising from the properties of the set system in NO instances). In addition, we have the constraint $\sum_{i=1}^k (x_i + y_i) \leq k$, meaning that we cannot allocate more than the total number of elements, $ks$. The following technical lemma bounds the value that can be obtained under these constraints. (We note that the parameters $\alpha, \beta$ in the lemma are related to $a, b$ by $\alpha = a \cdot s$ and $\beta = b/s$.)

**Lemma 4.3.** *Let $x_1, y_1, \ldots, x_k, y_k \geq 0$ be such that*

- *$x_1 \geq x_2 \geq \ldots \geq x_k$,*
- *for each $\ell \leq k$, $\sum_{i=1}^\ell x_i \leq (1 - (1 - 1/k)^\ell + \epsilon)k$,*
- *$\sum_{i=1}^k (x_i + y_i) \leq k$.*

*Let $v(x, y) = 1 - (1 - \alpha(2x - \beta))_+ (1 - \alpha(2y + \beta))_+$ if $x - \beta > y$, and $v(x, y) = 1 - (1 - \alpha(x + y))_+^2$ if $x - \beta \leq y$. Then,*

$$\sum_{i=1}^k v(x_i, y_i) \leq \max_{1 \leq k^* \leq k} 2\alpha k + 1 - \frac{\alpha^2 k^2}{k - k^*}((1 - 1/k)^{k^*} - \epsilon)^2$$

*Moreover, for $k = 2$ and $\beta \geq \frac{1}{4} + \epsilon$ we have that*

$$\sum_{i=1}^{2} v(x_i, y_i) = 2 - (1 - \alpha(2 - \frac{1 + \alpha\beta}{2\alpha}))_+^2.$$

We defer the proof to Appendix A. Given this lemma, we obtain bounds on the ratio between YES and NO instances.

**Corollary 4.4.** *The ratio between the social welfare in a NO instance and a YES instance is:*

- $1 - \frac{1}{2e} + O(\epsilon)$ for $\alpha = \frac{1}{2}, \beta = \epsilon$ and large enough $k$.
- $\frac{17}{18} + O(\epsilon)$ for $\alpha = \frac{2}{3}, \beta = \frac{1}{2} + 2\epsilon$ and $k = 2$.

*Proof.* We show the first part by optimizing for the best value of $k^*$ in the NO case: Let $z = k^*/k$. For large $k$, we can approximate $(1 - 1/k)^{2k^*} \simeq e^{-2z}$. Hence we maximize the function $\phi(z) = k - \frac{k}{4(1-z)}e^{-2z}$ over $z \in (0, 1)$ (the rest of the terms approach 0 and we ignore them in this analysis). By elementary calculus, $\phi(z)$ is maximized at $z = 1/2$, and the maximum value is $\phi(\frac{1}{2}) = k(1 - \frac{1}{2e})$. In other words, $k^* = k/2$ is the best choice, and gives

$$\sum_{i=1}^{k} v(x_i, y_i) = (1 - e^{-1/2})k + \frac{1}{2}k(1 - (1 - e^{-1/2})^2) = (1 - \frac{1}{2e})k.$$

In the YES case, the optimum is $k(1 - (1 - \alpha(2 - \beta))_+(1 - \alpha\beta)_+) = k(1 - \frac{1}{2}\epsilon(1 - \frac{1}{2}\epsilon)) \geq k(1 - \epsilon)$. Therefore, the ratio is $1 - \frac{1}{2e} + O(\epsilon)$.

For the second part, we choose $\alpha = \frac{2}{3}$ and $\beta = \frac{1}{2} + 2\epsilon$, which gives an optimal solution in the NO case $x_1 = 1 + \epsilon, y_1 = 0, x_2 = 0, y_2 = 1 - \epsilon$. The value of this solution is

$$\sum_{i=1}^{2} v(x_i, y_i) = 1 + 1 - (1 - \alpha y_2)_+^2 = 2 - (1 - \frac{2}{3}(1 - \epsilon))_+^2 = 2 - (\frac{1}{3} + \frac{2}{3}\epsilon)^2 \leq \frac{17}{9}.$$

In the YES case, as we argued, the optimum is $2(1 - (1 - \alpha(2 - \beta))_+(1 - \alpha\beta)_+) = 2(1 - (1 - \frac{2}{3}(2 - \frac{1}{2} - 2\epsilon))(1 - \frac{2}{3}(\frac{1}{2} + 2\epsilon))) = 2(1 - \frac{1}{3}\epsilon(\frac{2}{3} - \frac{1}{3}\epsilon)) \geq 2 - \epsilon$. Therefore, the ratio is $\frac{17}{18} + O(\epsilon)$. □

## 4.1 Communication complexity of welfare maximization

Here we prove the first part of Theorem 1.1, and Theorem 1.2: for every constant $\epsilon > 0$, a $(1 - \frac{1}{2e} + \epsilon)$-approximation requires exponential communication. For 2 players, a $(\frac{17}{18} + \epsilon)$-approximation requires exponential communication. This is the first constant-factor impossibility result for submodular valuations in the communication model. The basic idea is to reduce from the Set Disjointness problem. In the Set Disjointness problem there are $k$ players, each player $i$ holds a string $x^i \in \{0, 1\}^t$. The following theorem establishes the hardness of Set Disjointness.

**Theorem 4.5** ([1]). *It requires $\frac{t}{k^4}$ bits to distinguish between the following two cases:*

1. *There exists some $j$ such that $x_j^1 = \ldots = x_j^k = 1$.*

2. *For each two different players $i$ and $i'$ and every $j$ it holds that if $x_j^i = 1$ then $x_j^{i'} = 0$.*

*The lower bound holds also for randomized and non-deterministic algorithms.*

The plan is to reduce Set Disjointness (with an exponential $t$) to an instance of combinatorial auctions with multiple-peak submodular functions. Then we would like to show that a good approximation algorithm will let us decide the Set Disjointness problem. This will show that a good approximation algorithm requires exponential communication. The first step is to show the existence of an exponential well-structured set system.

**Lemma 4.6.** *For a universe $M$, $|M| = ks = m$ and for any $\epsilon > 0$, there is a collection of $t = 2^{\Theta(\epsilon^2 m/k^3)}$ partitions $\mathcal{P}^j = (C_1^j, \ldots, C_k^j)$ such that*

- *For each $j \in [t]$, $\bigcup_{i=1}^k C_i^j = M$.*
- *For each $i \in [k], j \in [t]$, $|C_i^j| = s$.*
- *For any $i \leq i'$ and any $j, j'$, $|C_i^j \cap C_{i'}^{j'}| \leq \frac{1+\epsilon}{k} s$.*
- *For every distinct $i_1, \ldots i_\ell$ and any $j_1, \ldots, j_\ell$, we have $|C_{i_1}^{j_1} \cup \ldots \cup C_{i_\ell}^{j_\ell}| \leq (1 - (1 - \frac{1}{k})^\ell + \epsilon)ks$.*

*Proof.* This is a slight modification of the probabilistic construction of [11]. Consider $M$ partitioned into $s$ $k$-tuples $(e_1^r, \ldots, e_k^r)$, $1 \leq r \leq s$. We generate each partition $\mathcal{P}^j$ as follows: from each $k$-tuple $(e_1^r, \ldots, e_k^r)$, we include each element in exacly one of the sets $C_1^j, \ldots, C_k^j$, using an independently random bijection $\{e_1^r, \ldots, e_k^r\} \to \{C_1^j, \ldots, C_k^j\}$. Thus the first two conditions are satisfied by construction.

Consider any pair $C_i^j, C_{i'}^{j'}$ for $i \neq i'$. Each set is generated by taking (independently) a random element from each $k$-tuple $(e_1^r, \ldots, e_k^r)$. Therefore, the probability of taking the same element is $1/k$ for each $k$-tuple and these events are independent. The expected size of the intersection $|C_i^j \cap C_{i'}^{j'}|$ is $s/k$, and by the Chernoff bound, the probability that the intersection is larger than $\Pr[|C_i^j \cap C_{i'}^{j'}| > (1+\epsilon)s/k] < e^{-\Omega(\epsilon^2 s/k)}$.

Similarly, consider any choice of distinct $(i_1, \ldots, i_\ell)$ and any $(j_1, \ldots, j_\ell)$. Each element appears in $C_{i_1}^{j_1} \cup \ldots \cup C_{i_\ell}^{j_\ell}$ with probability $1 - (1 - 1/k)^\ell$, and hence the expected cardinality of the union is $(1 - (1 - 1/k)^\ell)ks$. We can write $|C_{i_1}^{j_1} \cup \ldots \cup C_{i_\ell}^{j_\ell}| = \sum_{r=1}^s X_r$ where $X_r$ is the number of elements of the $r$-th $k$-tuple that are contained in the union. The random variables $X_r$ are independent, in the range $[0, k]$, and hence by the Chernoff bound $\Pr[\sum_{r=1}^s X_r > (1+\epsilon) \mathbf{E}[\sum_{r=1}^s X_r]] < e^{-\Omega(\epsilon^2 s/k)}$.

If the number of partitions is $t$, we have $O(t^k)$ combinations of sets to consider. By the union bound, the probability that any of the bad events above occurs is $O(t^k) e^{-\Omega(\epsilon^2 s/k)}$. For $t = 2^{\Theta(\epsilon^2 s/k^2)}$, the probability is smaller than 1 and hence there is a collection of partitions satisfying the assumptions. □

Given an instance of Set Disjointness, consider the following well structured collection of sets: We construct $C_i^j$ using Lemma 4.6 and we include $C_i^j$ in $\mathcal{S}_i$ if and only if $x_j^i = 1$. Consider the $k$ multiple-peak submodular functions that are based on this well structured collection of sets (with $b = \frac{1+\epsilon}{k} s$). Observe that if there exists some $j$ such that $x_j^1 = \ldots = x_j^k = 1$ then this is a YES instance. However, if for each two different players $i$ and $i'$ and every $j$ it holds that if $x_j^i = 1$ then $x_j^{i'} = 0$ this is a NO instance. Hence it requires exponential amount of communication to distinguish between these two cases (even for 2 players) and the hardness of approximation in the communication model follows from Corollary 4.4.

## 4.2 Computational complexity of welfare maximization

We now prove the second part of Theorem 1.1. The starting point of our reduction is Feige's inpproximability result for Max $k$-cover [14]. We summarize it here, with some additional useful properties of the set system that arises from Feige's reduction (see also [15] and [16] for some comments and explanations).

**Hardness of Max $k$-cover.** *For any fixed $\epsilon > 0$, it is NP-hard to distinguish the following two cases for a given collection of sets $\mathcal{S} \subset 2^M$, partitioned into groups $\mathcal{S}_1, \ldots, \mathcal{S}_k$:*

1. *YES case: There exist $k$ disjoint sets, one from each group $\mathcal{S}_i$, whose union is the universe $M$.*
2. *NO case: For any choice of $\ell \leq k$ sets, their union covers at most a $(1 - (1 - 1/k)^\ell + \epsilon)$-fraction of $M$.*

*In addition, the set system can be assumed to have the following properties:*

- *Every set has the same (constant) size $s$.*

- *Each group contains the same (constant) number of sets g.*

- *Every element appears in the same (constant) number of sets d.*

- *Any two sets intersect in at most $\epsilon s$ elements.*

We remark that the properties up to the last two are discussed in [15, 16]. The last two properties follow from the deterministic construction for max $k$-cover in [14], as follows:

The sets are unions of $(L-1)$-dimensional layers in disjoint hypercubes $[k']^L$, where $k'$ is the number of provers, and each hypercube corresponds to a particular random string $r$ generated by the verifier. (Here, $k'$ and $L$ are both constant.) Each set corresponds to a triple $(q, a, i)$ where $q$ is a question, $a$ is an answer, and $i$ is a prover. The direction of the layer that a set uses in a certain hypercube corresponds to the answer $a$, and the position (shift) of the layer corresponds to the prover $i$. Given a random string $r$ and a prover $i$, the question $q$ is determined by $r$ and $i$. Therefore different sets for the same prover $i$, restricted to the same hypercube $r$, must differ in their answer $a$. Hence, they are orthogonal in the respective hypercube. In particular, for each prover $i$ there can be at most $L$ different sets participating in the same hypercube, one for each dimension. As the number of provers and $L$ are constant, there is only a constant number of sets participating in each hypercube. As shown in [15], it can be arranged that the number of sets containing any given element is the same, and by the above it must be a constant.

For two different provers, the respective sets restricted to one hypercube are layers indexed by a different $i$, and hence they are either orthogonal or disjoint. Either way, any two different sets are either disjoint or intersect in a $1/k'$-fraction of their size in each hypercube where they both participate. Overall, two different sets can intersect in at most a $1/k'$-fraction of their elements. The number of provers $k'$ can be chosen large enough so that $1/k' \leq \epsilon$ (in any case $k'$ has to increase as $\epsilon \to 0$, for other reasons).

**Reduction.** We describe a reduction from the Max $k$-cover problem to a combinatorial auction with submodular valuations which works as follows. Given an instance of Max $k$-cover as above, we identify the $k$ groups of sets with $k$ players. The set of items is the universe $M$. For each player $i$, we define a multiple-peak submodular valuation function $v_i : 2^M \to \mathbb{R}_+$. The size of each set in $\mathcal{S}_i$ is $s$ (a constant), and the number of sets is $g$, also a constant. Hence, the set system $\mathcal{S}_i$ covers only a constant number of elements $\leq sg$, and we apply the construction of Definition 3.4 only to these elements (i.e. other elements have value 0 for player $i$). From the properties of the Max $k$-cover instance, we know that any two sets in $\mathcal{S}_i$ can overlap in at most $\epsilon s$ elements, hence the parameter $b$ can be set to $b = \epsilon s$. We also set $a = \frac{1}{2s}$. From the hardness of Max $k$-cover it follows now that it is NP-hard to distinguish between the YES and NO instances, with a ratio of optimal values $1 - \frac{1}{2e} + O(\epsilon)$ due to Corollary 4.4.

### 4.3 Max-min allocation

We note that the same proofs apply to the Max-min Allocation problem as well. This is because the reduction produces instances such that in the YES case, all players receive the same value. In the NO case, the minimum value is upper-bounded by the average value over all players. Therefore, the gap in terms of the objective $\min_{1 \leq i \leq k} v_i(S_i)$ is at least as large as in terms of welfare maximization. In fact the hardness factors can be (slightly) improved for Max-min allocation, but we defer this to the full version of the paper.

**Acknowledgment.** JV wishes to thank Uri Feige, who inspired the computational-complexity part of this paper in a conversation 6 years ago. We would also like to thank Ashwinkumar Badanidiyuru Varadaraja for valuable commments.

# A  Proof of Lemma 4.3

This section is devoted to proving Lemma 4.3. We consider some optimal solution and massage it to get the specified bounds. The following simple claims show several structural properties that an optimal solution can be assumed to possess.

**Claim A.1.** *The sequence $x_1 \geq x_2 \geq \ldots \geq x_k$ is non-zero up to a certain index $k^*$, $x_i - \beta > y_i$ for $i \leq k^*$ and then $x_i = 0$ for $i > k^*$.*

*Proof.* If $x_i - \beta \leq y_i$, we can assume that $x_i = 0$ (by replacing $y_i$ by $x_i + y_i$ and $x_i$ by 0, nothing changes). □

**Claim A.2.** *If $x_i = x_j$, $i < j \leq k^*$ then $y_i = y_j = 0$.*

*Proof.* Suppose that $x_i = x_j$, $i < j \leq k^*$ and suppose $y_j \leq y_i \neq 0$. The derivative $\frac{\partial v}{\partial y}$ at $x_i$ equals the derivative at $x_k$. So we can move some small amount of mass from $y_i$ to $y_j$, while making sure that still $x_j - \beta > y_j$. Observe that the value of the solution is the same. But once $y_i > y_j$, it is already more profitable to increase $x_i$. So the case where $x_i = x_j$ is not optimal, unless $y_i = y_j = 0$. □

**Claim A.3.** *All the variables $y_i$ for $i > k^*$ are equal. We denote their value by $y^*$.*

*Proof.* If not, we could gain by making them equal, due to the concavity of $v(0,y) = 1 - (1 - \frac{1}{2}y)_+^2$. □

**Claim A.4.** *For all $y_i < k^*$, $y_i = 0$.*

*Proof.* Consider some $y_i < k^*$ with $y_i > 0$. Notice that $x_i > x_{k^*}$ (by Claim A.2 if $x_i = x_{k^*}$ then $y_i = 0$), and hence the derivative $\frac{\partial v}{\partial y}$ at $x_i$ is smaller than the derivative $\frac{\partial v}{\partial y}$ at $x_{k^*}$. So moving some mass from $y_i$ to $y_{k^*}$ improves the value of the optimal solution — a contradiction. □

## A.1  The case of 2 players

**Claim A.5.** *$k^* \geq 1$.*

*Proof.* Suppose $k^* = 0$. Then, by the previous claims we have that $x_1 = x_2 = 0$ and $y_1 = y_2 = 1$. The value of this solution is $2 - 2(1-\alpha)_+^2$. However the solution in which $x_1 = y_2 = 1$ and $x_2 = y_1 = 0$ has value $2 - (1-\alpha)_+^2 - (1-\alpha(2-\beta))_+(1-\alpha\beta)_+$ which is no smaller by using simple arithmetic. □

**Claim A.6.** *$y_1 = 0$.*

*Proof.* Suppose $y_1 > 0$. It must be that $k^* = 1$ (by Claim A.4). If $1 - a(2x_1 - b) \leq 0$ then $y_1$ does not bring any benefit and moving all the mass of $y_1$ to $y_2$ can only improve the value of the solution. Otherwise, moving some mass from $y_1$ to $x_1$ (making sure that still $1 - a(2x_1 - b) \geq 0$) will improve the optimal solution — a contradiction. □

**Claim A.7.** *If $\beta \geq \frac{1}{2} + 2\epsilon$ then $k^* = 1$.*

*Proof.* By the constraints of the lemma we have that $x_1 + x_2 + y_1 + y_2 \leq 2$ and $x_1 + x_2 \leq 2(\frac{3}{4} + \epsilon)$. Without loss of generality, $x_1 + x_2 + y_1 + y_2 = 2$, and hence $x_1 + x_2 - y_1 - y_2 = 2(x_1 + x_2) - (x_1 + x_2 + y_1 + y_2) \leq 1 + 4\epsilon$. But if $k^* = 2$, summing up $x_1 - \beta > y_1$ and $x_2 - \beta > y_2$ we get $x_1 + x_2 - y_1 - y_2 > 2\beta \geq 1 + 4\epsilon$. This is a contradiction. □

**Claim A.8.** $x_1 = \frac{1+\alpha\beta}{2\alpha}$.

*Proof.* $x_1 = \frac{1+\alpha\beta}{2\alpha}$ is the point at which $1 - \alpha(2x_1 - \beta) = 0$, i.e. the first player is saturated and does not benefit from any additional items. If $1 - \alpha(2x_1 - \beta) < 0$, we can move some mass from $x_1$ to $y_2$ and improve the optimal solution. If $1 - \alpha(2x_1 - \beta) > 0$, we can move some mass from $x_2$ or $y_2$ to $x_1$ and improve the optimal solution. □

This implies that the following is an optimal solution: $x_1 = \frac{1+\alpha\beta}{2\alpha}$, $y_2 = 2 - \frac{1+\alpha\beta}{2\alpha}$, $x_2 = y_1 = 0$. The first player gets value 1, and the second player $1 - (1 - \alpha y_2)_+^2$; hence the optimum is

$$\sum_{i=1}^{2} v(x_i, y_i) = 1 + 1 - (1 - \alpha y_2)_+^2 = 2 - (1 - \alpha(2 - \frac{1+\alpha\beta}{2\alpha}))_+^2.$$

This proves the second part of Lemma 4.3.

## A.2 The case of many players

By the series of claims we proved above, we have a sequence $x_1 \geq x_2 \geq \ldots \geq x_{k^*}$, $y_i = 0$ for $i < k^*$, and then $y_{k^*+1} = y_{k^*+2} = \ldots = y_k = y^*$, $x_i = 0$ for $i > k^*$. The value of this solution is at most (we upper bound the value of player $k^*$ by 1):

$$\sum_{i=1}^{k} v(x_i, y_i) = \sum_{i=1}^{k^*-1} (1 - (1 - \alpha(2x_i - \beta))_+(1 - \alpha\beta)_+) + 1 + \sum_{i=k^*+1}^{k} (1 - (1 - \alpha y^*)_+^2)$$

$$\leq \sum_{i=1}^{k^*-1} 2\alpha x_i + 1 + \sum_{i=k^*+1}^{k} (1 - (1 - \alpha y^*)_+^2).$$

We have the global constraint $\sum_{i=1}^{k^*} x_i + \sum_{i=k^*+1}^{k} y_i \leq k$, and also $\sum_{i=1}^{k^*} x_i \leq k(1 - (1 - 1/k)^{k^*} + \epsilon)$. Subject to these constraints, it is profitable to make the variables $x_i$ as large as possible, because they bring more benefit than the $y_i$ variables (see the formula above). Therefore, we can assume that both constraints are tight: $\sum_{i=1}^{k^*} x_i = k(1 - (1 - 1/k)^{k^*} + \epsilon)$, and $\sum_{i=k^*+1}^{k} y_i = (k - k^*)y^* = ((1 - 1/k)^{k^*} - \epsilon)k$. Then we have $y^* = \frac{k}{k-k^*}((1 - 1/k)^{k^*} - \epsilon)$. The objective value is

$$\sum_{i=1}^{k} v(x_i, y_i) \leq 2\alpha \sum_{i=1}^{k^*-1} x_i + 1 + (k - k^*)(1 - (1 - \alpha y^*)^2)$$

$$= 2\alpha k(1 - (1 - \frac{1}{k})^{k^*} + \epsilon) + 1 + (k - k^*)(1 - (1 - \frac{\alpha k}{k - k^*}((1 - 1/k)^{k^*} - \epsilon))^2)$$

$$= 2\alpha k(1 - (1 - \frac{1}{k})^{k^*} + \epsilon) + 1 + (k - k^*)(\frac{2\alpha k}{k - k^*}((1 - 1/k)^{k^*} - \epsilon) - \frac{\alpha^2 k^2}{(k - k^*)^2}((1 - 1/k)^{k^*} - \epsilon)^2)$$

$$= 2\alpha k + 1 - \frac{\alpha^2 k^2}{k - k^*}((1 - 1/k)^{k^*} - \epsilon)^2.$$

This proves the first part of Lemma 4.3.

# B Combinatorial Public Projects

Here, we prove Theorem 1.3, in two parts. Due to the similarities with welfare maximization, we only point out the necessary differences.

## B.1 Communication complexity of CPP

We reduce from an instance of Set Disjointness to CPP with submodular valuations as follows. We use Lemma 4.6 to construct a collection of partitions $(C_1^j, \ldots, C_k^j), 1 \leq j \leq t$. The difference now is that we use $C_1^j$ for all $k$ players, rather than $C_i^j$ for player $i$. We include $C_1^j$ in collection $\mathcal{S}_i$ if $x_j^i = 1$ in the Set Disjointness instance. By Lemma 4.6, $\mathcal{S}_i$ is a well structured collection. The valuation $v_i$ of player $i$ is going to be the $(\mathcal{S}_i, a, b)$-multiple-peak submodular function of Definition 3.4, where $a = \frac{1}{2s}$ and $b = \frac{1+\epsilon}{k}s$. The CPP instance is going to be $\max\{\sum_{i=1}^k v_i(S) : |S| = s\}$.

In the YES case, all players share an index $j$ such that $x_j^1 = x_j^2 = \ldots = x_j^k = 1$, and hence $C_1^j \in \mathcal{S}_i$ for all $i \in [k]$. Then $C_1^j$ is a feasible solution of value $v_i(S_1^j) = 1 - (1 - a(2s - b))_+(1 - ab)_+ = 1 - \frac{1+\epsilon}{2k}(1 - \frac{1+\epsilon}{2k}) = 1 - O(1/k)$ for each player.

In the NO case, for every index $j$ there is at most one player $i$ such that $x_j^i = 1$; hence $C_1^j \in \mathcal{S}_i$ for at most one player $i$. By Lemma 4.6, $|C_1^j \cap C_1^{j'}| \leq \frac{1+\epsilon}{k}s = b$. By Proposition 3.2, any set $S$ can be $b$-close to at most one set $C_1^j$. Therefore, for any feasible set $S$, at most one player derives a value defined by a $b$-close set, and all the remaining $k-1$ players get value defined by the formula $v_i(S) = 1 - (1 - a|S|)_+^2 = 1 - (1 - as)_+^2 = 3/4$. One player can receive value close to 1.

For a large number of players ($k \to \infty$), the value per player tends to 1 in the YES case and $3/4$ in the NO case. Therefore, if we could approximate the CPP problem within a factor better than $3/4$, we could solve the Set Disjointness problem.

For 2 players, we still use the construction above, but we only consider the first two players. In the YES case, both players get value close to 1. In the NO case, 1 player must get value close to $3/4$, while the other player can get value close to 1. Thus if we could approximate the CPP problem for 2 players within a factor better than $7/8$, we could solve the Set Disjointness problem.

## B.2 Computational complexity of CPP

Here we prove the second part of Theorem 1.3. This is actually quite simple and does not require our construction of multiple-peak submodular functions. The same reduction has been used in [6] to prove that CPP is hard to approximation better than $1 - 1/e$ with unit-demand valuations. The only new ingredient here is the observation that the valuation functions can be assumed to depend only on a constant number of items. The result of [6] already implies that the factor of $1 - 1/e$ for CPP with submodular valuations cannot be improved in the demand oracle model; we strengthen this to the case of constant-size valuation functions.

*Proof of Theorem 1.3, part 2.* Consider an instance of Max $k$-cover as above. We associate players with the elements of the universe $M$, and items with the sets in $\mathcal{S}$. For each player $e \in M$, we define a valuation function as follows: $v_e(T) = 1$ if $T$ contains any set covering element $e$, and 0 otherwise. Due to the properties of the set system, this valuation function depends non-trivially only on a constant number of items ($d$), and it is a very special monotone submodular function (unit demand) of the form $v_e(T) = \min\{|T \cap D_e|, 1\}$.

The objective function of the CPP problem is now $\sum_{e \in M} v_e(T)$ which is the number of elements covered by the sets in $T$. Subject to the condition $|T| \leq k$, this is NP-hard to maximize within a $1 - 1/e + \epsilon$ factor. □